# Predicting fall risk in older adults: A machine learning comparison of accelerometric and non-accelerometric factors

Ana González-Castro[1] , José Alberto Benítez-Andrades[2] , Rubén González-González[3], Camino Prada-García[4,5] and Raquel Leirós-Rodríguez[6]

## Abstract

**Objectives:** Accurate prediction of fall risk in older adults is essential to prevent injuries and improve quality of life. This study evaluates the predictive performance of various machine learning models using accelerometric data, non-accelerometric data, aiming to improve predictive accuracy and identify key contributing variable.
**Methods:** We applied random forest, XGBoost, AdaBoost, LightGBM, support vector regression (SVR), decision trees, and Bayesian ridge regression to a dataset of 146 older adults. Models were trained using accelerometric data (movement patterns) and non-accelerometric data (demographic and clinical variables). Performance was evaluated based on mean squared error (MSE) and coefficient of determination ($R^2$), to assess how combining multiple data types influences prediction accuracy.
**Results:** Models trained on combined accelerometric and non-accelerometric data consistently outperformed those based on single data types. Bayesian ridge regression achieved the highest accuracy (MSE = 0.6746, $R^2$ = 0.9941), demonstrating superior performance compared to decision trees (MSE = 0.1907, $R^2$ = 0.8991) and SVR (MSE = 1.5243, $R^2$ = −2.2532). Non-accelerometric factors, including age and comorbidities, significantly contributed to fall risk prediction.
**Conclusions:** Integrating accelerometric and non-accelerometric data improves fall risk prediction accuracy in older adults. Bayesian ridge regression trained on combined datasets provides superior predictive power compared to traditional models. These findings highlight the importance of multi-source data fusion for effective fall prevention strategies. Future work should validate these models in larger, more diverse populations to enhance clinical applicability.

[1]Nursing and Physical Therapy Department, Universidad de León, Ponferrada, Spain
[2]SALBIS Research Group, Department of Electric, Systems and Automatics Engineering, Universidad de León, León, Spain
[3]Department of Electric, Systems and Automatics Engineering, Escuela de Ingenierías Industrial, Informática y Aeroespacial, Universidad de León, León, Spain
[4]Department of Preventive Medicine and Public Health, University of Valladolid, Valladolid, Spain
[5]Dermatology Service, Complejo Asistencial Universitario de León, León, Spain
[6]SALBIS Research Group, Nursing and Physical Therapy Department, Universidad de León, Ponferrada, Spain

**Corresponding author:**
Camino Prada-García, Department of Preventive Medicine and Public Health, University of Valladolid, 47005 Valladolid, Spain; Dermatology Service, Complejo Asistencial Universitario de León, 24008 León, Spain.
Email: cprada@saludcastillayleon.es

## Introduction and related work

### Background on fall risk

Falls among older adults are a major health concern, with one-third experiencing falls annually, and up to 20% resulting in serious injuries such as fractures or head trauma.[1,2] This problem is compounded by an aging population and places a significant economic burden on healthcare systems, exceeding 2 billion dollars annually in countries like Canada.[3]

Beyond physical injuries, falls reduce functional independence and quality of life.[4] They often lead to prolonged hospitalizations, institutionalization, and increased mortality.[5] Additionally, the fear of falling can discourage physical activity, creating a cycle of physical decline that further elevates fall risk.[6]

The financial burden of falls is expected to increase as populations age, reinforcing the urgent need for effective fall prevention and improved risk prediction methods to mitigate both health and economic consequences.

Given the complexity of fall risk factors—including mobility impairments and chronic illnesses—Hopewell et al.[7] and LaPorta et al.[8] emphasized the necessity of a multidimensional approach to effectively address this issue. Montero-Odasso et al.[9] further suggested that personalized interventions based on the comprehensive risk assessment are more effective in fall prevention than unidimensional approaches.

### Role of accelerometric data

Fall risk prediction has been a major focus in both medical and technological research. Pooranawatthanakul and Siriphorn,[10] as well as Urbanek et al.,[11] demonstrated that accelerometric data from wearable sensors are valuable for identifying abnormal movement patterns that precede falls. These sensors provide real-time insights into gait stability, cadence, and movement variability, essential for assessing fall risk. Schootemeijer et al.[12] reported that wearable accelerometers outperform traditional clinical assessments in predicting falls by monitoring balance and gait speed.

Recent studies have further explored deep learning methods for fall risk prediction, demonstrating that neural networks can enhance predictive accuracy by capturing complex movement patterns.[13]

While wearable sensors offer valuable insights, their use in fall risk prediction presents practical limitations. Sensor placement variability—whether worn on the wrist, waist, or ankle—can significantly impact the accuracy of extracted features.[11] Additionally, user compliance remains a major challenge, as older adults may forget to wear the device, misposition it, or remove it due to discomfort, leading to missing or inconsistent data.[12]

These limitations highlight the need for multi-source data fusion, where accelerometric information is complemented by clinically relevant non-accelerometric variables such as age, medical history, and functional assessments. Integrating multiple data sources reduces dependency on sensor adherence and enhances the robustness of fall risk models.

### Need for integration

However, accelerometric data alone do not capture the full complexity of fall risk. Non-accelerometric factors, such as age and chronic conditions, play a crucial role in fall susceptibility.[14] Studies have shown that combining both data types improves model accuracy, allowing for a more comprehensive risk assessment. Lien et al.[15] demonstrated that this integration is particularly beneficial for detecting subtle gait and balance changes that may not be evident through movement data alone. Thus, while accelerometric data provide key movement insights, incorporating non-accelerometric factors results in a more holistic and clinically relevant fall risk evaluation.

Recent studies have explored advanced data processing techniques to optimize wearable sensor data for fall risk assessment. For instance, alternative sampling and augmentation methods have been proposed to enhance the reliability of deep learning models in analyzing movement patterns.[16] These strategies emphasize the need for refined preprocessing approaches to improve predictive performance.

While accelerometric data are valuable for fall risk prediction, they alone may not capture the full range of contributing factors. Non-accelerometric data, such as age, comorbidities, and environmental influences, provide critical context. Urbanek et al.[11] and Thiamwong et al.[17] demonstrated that while accelerometry-based assessments effectively detect movement-related risks, they may overlook key clinical and demographic risk factors.

Moreover, integrating accelerometric and non-accelerometric data enhances the accuracy of fall risk prediction models. Schootemeijer et al.[12] and Lien et al.[15] showed that models combining both data types outperform accelerometry-only approaches, particularly in differentiating risk levels and improving sensitivity and specificity. This integration enables a more comprehensive risk assessment.

Despite these advancements, many studies still face methodological limitations. Antonietti[18] and Millet et al.[19] focused solely on either accelerometric or non-accelerometric data, overlooking the benefits of combining them. Even when both data types are considered, many models rely on traditional statistical techniques, which lack the predictive power and flexibility of modern machine learning methods.[20,21]

Deep learning models have shown promise in fall detection, but their application to fall risk prediction—especially with diverse data sources—remains underexplored.[13] This study bridges this gap by using ensemble machine learning models to capture both movement patterns and clinical risk factors, improving predictive accuracy and interpretability.

To achieve this, we integrate heterogeneous data sources, apply advanced machine learning techniques, and utilize a robust dataset to enhance the predictive accuracy and generalizability of fall risk assessment models.

In summary, while accelerometric data are essential for detecting movement abnormalities, non-accelerometric data provide critical clinical context, enriching predictive models. A comparative analysis of these data types is crucial for developing robust and accurate fall risk prediction tools.

## Objectives of the study

The primary objective of this study is to evaluate and compare the effectiveness of various machine learning algorithms in predicting fall risk among older adults using accelerometric data, non-accelerometric data, and their combination. Specifically, this study aims to:

1. Compare the predictive accuracy of multiple machine learning models, including random forest, XGBoost, AdaBoost, LightGBM, Bayesian ridge, support vector regression (SVR), and decision trees, when applied to accelerometric, non-accelerometric, and combined datasets. This analysis will identify the strengths and weaknesses of each model across different data contexts.[18,19]
2. Determine which data type—accelerometric, non-accelerometric, or combined—provides the highest predictive performance based on key metrics such as accuracy, mean squared error (MSE), and the coefficient of determination ($R^2$).[20,21]
3. Analyze the relative importance of individual variables in predicting fall risk, identifying key accelerometric and non-accelerometric factors that contribute most to model performance.[22]
4. Assess the practical implications of the findings for fall prevention strategies, providing insights that can guide the development of predictive tools for clinical applications.[23]

Furthermore, this study utilizes ensemble learning methods such as XGBoost and LightGBM, widely adopted in structured data analysis for their ability to capture complex feature interactions while maintaining interpretability.[20] Unlike deep learning, which often requires large-scale datasets, ensemble models achieve high predictive performance with limited data, making them well-suited for fall risk assessment, where sample sizes are often constrained.

By integrating heterogeneous data sources and applying advanced machine learning techniques, this study contributes to the development of hybrid predictive models. Future research should explore combining deep learning with ensemble methods to further enhance fall risk prediction, optimizing accuracy, interpretability, and clinical applicability.

## Contribution of the study

The goal of this study is to provide evidence-based insights that enhance fall risk assessment, ensuring that predictive models are both accurate and practical for real-world applications in fall prevention.

This work advances fall risk prediction by comprehensively comparing machine learning algorithms that integrate accelerometric and non-accelerometric data. Unlike previous studies that focus on only one data type, this approach develops a more robust predictive model.[18,19] The findings underscore the importance of combining diverse data sources to improve model accuracy and reliability, addressing a key gap in the literature.[21,23]

Additionally, this study advances the application of machine learning in real-world settings by evaluating the performance of different algorithms, such as random forest, XGBoost, AdaBoost, and LightGBM, across various data configurations. The results provide valuable insights into the computational efficiency and practical applicability of these models in clinical environments, ultimately aiming to enhance fall prevention strategies and reduce fall-related incidents among older adults.[20,22]

Overall, this research contributes to developing more accurate and comprehensive fall risk assessment tools that can be utilized in healthcare settings to identify at-risk individuals better and implement targeted interventions.

The paper is structured as follows: the "Methodology" section delineates the machine learning models employed, details the hyperparameter adjustments, and elucidates the oversampling techniques utilized. It also provides the rationale behind the selection of specific variable groupings. The "Experiments and results" section offers a comprehensive description of the dataset and presents the experimental setup and findings. Finally, the "Discussion" section delves into the interpretation of the findings, their broader implications, and potential limitations. The "Conclusions" section summarizes the key takeaways of the research and outlines future research avenues and potential improvements.

## Methodology

The methodology used to carry out this research, from data collection to modeling results, is summarized in Figure 1.

## Dataset description and preprocessing

This study analyzed data from 146 older adults (aged 65 and above) collected over 12 months during routine health assessments at a community health center. The dataset includes both accelerometric data from wearable sensors and non-accelerometric data, such as demographic, psychometric, and clinical variables, providing a comprehensive assessment of fall risk factors.

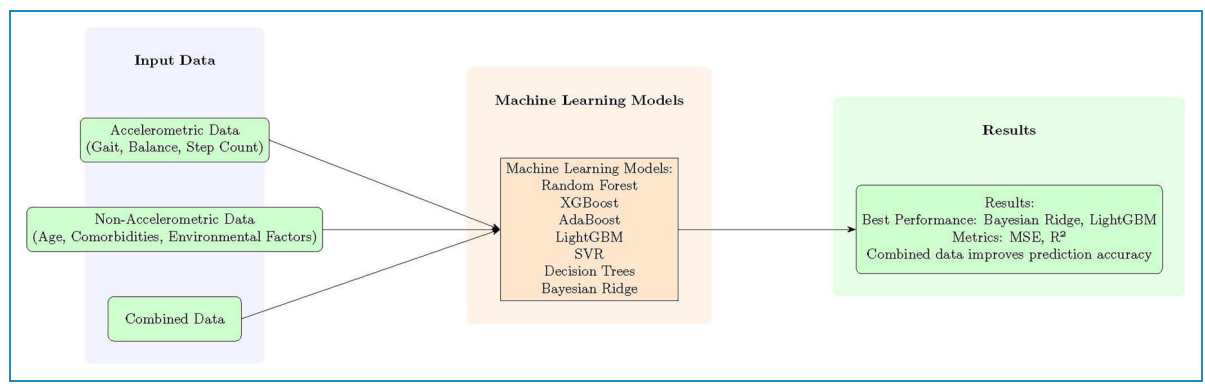

**Figure 1.** Methodology applied in research from data collection on older adults to the development of predictive models for fall risk.

The collected variables are categorized into two primary groups: accelerometric and non-accelerometric (Table 1). Wearable sensors continuously monitored participants' movements, capturing gait, balance, and overall physical activity for a more detailed fall risk evaluation.

Accelerometric data capture three key aspects: acceleration magnitude, the time at which it occurs, and the plane of motion. By combining these parameters, a comprehensive set of accelerometric variables is generated to analyze balance and motor control during gait and functional tests.

Examples include:

- Maximum and minimum accelerations in the vertical axis during gait.
- Time point where acceleration peaks in the vertical or medio-lateral axis during gait or functional tests.
- Maximum acceleration detected in the sagittal plane.
- Time point of minimum acceleration in the anterior-posterior axis.
- Root mean squared (RMS) acceleration as a measure of overall movement variability.

The complete set of accelerometric variables is available in Supplemental Table 1, while the variables analyzed in this study are listed in Table 1.

In addition to accelerometric data, the dataset includes a comprehensive set of non-accelerometric variables, encompassing intrinsic factors such as age, gender, and medical history (including chronic conditions like hypertension and diabetes). It also incorporates psychometric assessments to evaluate cognitive function and fear of falling.

By integrating both accelerometric and non-accelerometric variables, this study enables a multi-faceted analysis of fall risk, considering both physical and contextual factors that influence fall likelihood in older adults.

Before applying machine learning algorithms, the dataset underwent a rigorous preprocessing phase to ensure data quality and consistency.

Notably, no missing values were present, as all 337 variables were fully recorded for each of the 146 participants. Consequently, no imputation or interpolation techniques were required.

Variable selection was based on relevance in fall risk assessment literature and availability in our dataset, ensuring a comprehensive evaluation of contributing factors. Both accelerometric and non-accelerometric variables were included to capture movement-based and contextual risk factors.

Additionally, as all participants were women, no sex-related variability was present. The dataset was also assessed for class balance, confirming that fallers and non-fallers were adequately represented, reducing potential model bias.

Additionally, all variables were normalized or standardized depending on their distribution characteristics to ensure that they contribute equally to the model training process. Categorical variables, such as medical history and gender, were encoded using one-hot encoding to facilitate their inclusion in the models. Outliers, particularly in the accelerometric data, were identified and handled using a combination of statistical methods and domain expertise to minimize their impact on model performance. This preprocessing step was critical for preparing the data for effective model training and evaluation.

## Machine learning algorithms used

To predict fall risk, we employed seven machine learning algorithms: random forest, XGBoost, AdaBoost, LightGBM, Bayesian ridge, SVR, and decision trees. These algorithms were selected due to their robustness, scalability, and demonstrated effectiveness in handling complex datasets and performing well in predictive modeling tasks.[20,22] Given that our dataset integrates structured clinical variables (non-accelerometric) with accelerometric time-series data, it was crucial to employ models capable of handling both feature types and capturing non-linear relationships.

Table 1. Description of the variables selected and their possible values.

| Category | Variable | Values |
| --- | --- | --- |
| **Medical history and anthropometry** | | |
| AGE | Age of the participants | Continuous (years) |
| SP_TRUNK | Trunk span of the participants | Continuous (cm) |
| BMI | Body mass index | Continuous (kg/m$^2$) |
| HEART | Presence of heart disease or condition | 0 = no, 1 = yes |
| DEGENERATIVE | Presence of degenerative disease or condition | 0 = no, 1 = yes |
| **Functional tests** | | |
| TUG | Timed up & go test result | Continuous (s) |
| 6MWT | 6-minutes walking test result | Continuous (m) |
| S_LL | Lower limbs strength (Rikli & Jones test) | Continuous (repetitions) |
| FLX_LL_R | Right lower limb flexibility (Rikli & Jones test) | Continuous (cm) |
| FLX_LL_L | Left lower limb flexibility (Rikli & Jones test) | Continuous (cm) |
| FLX_UL_L | Left upper limb flexibility (Rikli & Jones test) | Continuous (cm) |
| **Body composition** | | |
| B_MET | Basal metabolic rate | Continuous (kcal/day) |
| FM_TRUNK | Trunk fat mass | Continuous (kg) |
| IMP_LL_R | Right lower limb impedance | Continuous ($\Omega$) |
| IMP_LL_L | Left lower limb impedance | Continuous ($\Omega$) |
| **Accelerometry of gait test** | | |
| MINseg_AX1 | Second of the test in which the minimum acceleration in the vertical axis during walking was identified | Continuous (n) |
| MAXseg_AX2 | Second of the test in which the maximum acceleration in the medio-lateral axis during walking was identified | Continuous (n) |
| MAX_RMS | Maximum acceleration detected in the mean root square of the three axes while gait | Continuous (G) |
| MINseg_$\alpha$ | Second of the test in which the minimum acceleration in the transverse plane during walking was identified | Continuous (n) |
| MAXseg_$\beta$ | Second of the test in which the maximum acceleration in the sagittal plane during walking was identified | Continuous (n) |

**Random forest** is an ensemble learning method that constructs multiple decision trees during training and outputs the mode of the classes for classification tasks. It is particularly advantageous due to its ability to handle heterogeneous data types, including categorical and numerical features, which aligns with the mixed nature of accelerometric and non-accelerometric variables in our dataset. Random forest is also known for its resilience to noisy data and its capability to rank the importance of features within the model, which is critical in understanding the key variables influencing fall risk.[19]

**XGBoost (extreme gradient boosting)** is a powerful ensemble method renowned for its speed and high performance. It builds on the gradient boosting framework by optimizing computational resources and reducing overfitting through regularization techniques. XGBoost was chosen specifically for its ability to model complex, non-linear interactions between accelerometric and clinical data, making it well-suited for fall risk prediction. XGBoost has consistently outperformed other models in both academic research and industry applications, particularly in scenarios where structured data and model interpretability are key considerations.[20,23]

**AdaBoost** (adaptive boosting) is an ensemble technique that sequentially adjusts the weights of misclassified instances, allowing the model to focus on the more challenging cases in subsequent iterations. This characteristic is particularly useful in datasets where subtle variations in movement patterns or clinical conditions contribute to fall risk, making AdaBoost a strong candidate for refining classification in these cases.[21]

**LightGBM (light gradient boosting machine)** is an efficient gradient boosting framework designed to handle large-scale data with high dimensionality. It achieves this by using a histogram-based approach to split data, resulting in faster training times and lower memory usage compared to traditional gradient boosting methods. Given the high dimensionality of the dataset (337 features), LightGBM was selected for its efficiency in handling a large number of input variables while maintaining predictive accuracy.[18]

**Bayesian ridge** is a linear regression model that incorporates a Bayesian approach, allowing for the regularization of parameters. This technique is particularly useful in avoiding overfitting, especially in cases where the number of features exceeds the number of observations. As our dataset consists of 146 participants but 337 variables, Bayesian Ridge was included to mitigate overfitting risks and to provide probabilistic insights into fall risk predictions.[24]

**Support vector regression (SVR)** is a regression technique based on the principles of support vector machines (SVMs). SVR is effective in high-dimensional spaces and can model non-linear relationships through the use of kernel functions. It was initially included due to its theoretical strengths in non-linear regression; however, its high computational cost and poor scalability to datasets with hundreds of variables resulted in suboptimal performance, as observed in our results.[25]

**Decision trees** are a non-parametric supervised learning method used for both classification and regression tasks. By creating a model that predicts the value of a target variable by learning simple decision rules inferred from the data features, decision trees are easy to interpret and useful for identifying the most significant variables in a dataset. However, they can be prone to overfitting, especially in the absence of pruning techniques or ensemble methods like random forest. In this study, decision trees were included as a baseline model to compare against more sophisticated approaches and to provide insight into feature importance.[26]

Each of these algorithms was chosen for its unique strengths, contributing to a comprehensive comparison of model performance in predicting fall risk among older adults.

### Model configuration

The machine learning models were configured and trained using a systematic approach to optimize performance. Hyperparameter tuning was conducted via grid search with five-fold cross-validation, testing multiple parameter combinations to identify the optimal configuration for each model.

The key hyperparameters adjusted for each algorithm were:

- **Random forest**: Number of trees (estimators) and maximum tree depth.
- **XGBoost**: Learning rate, maximum depth, and number of boosting rounds.
- **AdaBoost**: Number of estimators and learning rate.
- **LightGBM**: Number of leaves, learning rate, and boosting type.
- **Support vector regression (SVR)**: Kernel type, regularization parameter ($C$), and epsilon (controls margin width).
- **Decision trees**: Maximum depth and minimum samples required to split a node.
- **Bayesian ridge regression**: Regularization parameters (alpha and lambda) to control model complexity and prevent overfitting.

All models were trained on the processed dataset, with 80% of the data used for training and 20% for testing. To address class imbalance, oversampling techniques such as SMOTE (synthetic minority over-sampling technique) and RandomOverSampler were applied, ensuring that the minority class (high fall risk cases) was adequately represented. Model performance was evaluated on the test set to assess generalization capability.

Hyperparameter tuning was computationally intensive due to the diversity of models tested. Grid search was executed on an Intel Core i9 processor with 64 GB RAM, utilizing parallel processing to enhance efficiency.

The execution time varied by model:

- Tree-based ensemble methods (random forest, XGBoost, and LightGBM) required 4 to 6 hours due to their extensive hyperparameter search space.
- Bayesian ridge and SVR completed tuning in under an hour.
- SVR exhibited a particularly high computational cost relative to its performance, further reinforcing its limitations for high-dimensional datasets.

This computational analysis ensured that all models were optimized effectively while considering practical constraints for real-world applications.

This careful configuration and optimization process ensured that each model was trained under optimal conditions, maximizing their ability to accurately predict fall risk in older adults.

## Model evaluation metrics

To evaluate the performance of the machine learning models, several key metrics were employed, each offering insights into different aspects of the models' predictive capabilities. The primary metrics used in this study were MSE, mean absolute error (MAE), and the coefficient of determination ($R^2$). These metrics are widely recognized in the literature for their effectiveness in providing a balanced evaluation of model performance across various predictive tasks.[27,28]

**Mean squared error (MSE)** was employed to evaluate the average squared difference between the predicted and actual values, particularly in regression contexts. MSE is a critical metric because it penalizes larger errors more heavily, thereby providing a stringent assessment of the precision of the model's predictions. It is especially useful in identifying models that perform poorly due to large deviations in predictions.[29]

**Mean absolute error (MAE)** complements MSE by measuring the average absolute difference between predicted and actual values. Unlike MSE, which disproportionately penalizes large errors, MAE provides a more interpretable measure of model performance in the original units of the target variable. Including MAE allows for a more comprehensive evaluation of prediction errors, particularly in scenarios where extreme values might distort MSE results.

**Coefficient of determination ($R^2$)** quantifies the proportion of variance in the dependent variable that is predictable from the independent variables. An $R^2$ close to 1 indicates that the model explains a large portion of the variance, suggesting a good fit, whereas an $R^2$ close to 0 implies that the model does not capture the variance well, highlighting potential shortcomings in the model's explanatory power.[19]

*Statistical validation and confidence intervals (CIs).* To ensure the robustness of our findings, we computed 95% CIs for MSE, MAE, and $R^2$ using a bootstrapping approach with 1000 resamples. This statistical validation accounts for variability in the dataset, providing a more reliable assessment of model performance.

Given that our study focuses on a regression problem, classification-specific metrics such as F1-score and confusion matrices are not applicable in this context. Instead, our evaluation prioritizes metrics that effectively quantify prediction error and explained variance, ensuring an accurate representation of model performance in fall risk prediction.

## Comparison of models and variable importance analysis

The performance of the machine learning models was systematically compared across three datasets: accelerometric data, non-accelerometric data, and a combination of both. Each model was evaluated using MSE and the coefficient of determination ($R^2$) to identify the best-performing algorithm and dataset combination.

To facilitate direct comparisons, the results were presented in both tabular and graphical formats. MSE was used to assess overall prediction error, while $R^2$ provided insights into explanatory power. The analysis highlighted the strengths and weaknesses of each algorithm, with particular focus on their performance using the combined dataset.

Models trained on the combined dataset consistently outperformed those using only accelerometric or non-accelerometric data. Among the algorithms, Bayesian ridge regression and LightGBM demonstrated superior performance, reinforcing the benefit of integrating diverse data types. These findings emphasize the importance of a comprehensive data-driven approach to enhance fall risk prediction in older adults.

To assess the contribution of each variable to the predictive models, a variable importance analysis was performed. For ensemble models (random forest and LightGBM), feature importance was calculated using Gini impurity (random forest), and split gain (LightGBM), which measure each variable's contribution to reducing prediction error.

For non-ensemble models (SVR and Bayesian ridge regression), SHAP (SHapley Additive exPlanations) values were used to quantify each variable's impact on the model's predictions. SHAP values provide a consistent framework for feature importance, offering greater interpretability, particularly in complex models.

The analysis confirmed that both accelerometric and non-accelerometric variables play critical roles in predicting fall risk. Among accelerometric features, gait stability and step count were the most influential, while age and history of previous falls emerged as key non-accelerometric factors. These findings highlight the complementary nature of both data types, reinforcing the value of integrated models for robust fall risk prediction.

### Ethical considerations and study limitations

This study was conducted in accordance with the ethical standards set by the Institutional Review Board (IRB) overseeing the research. Informed consent was obtained from all participants before data collection, ensuring they were fully aware of the study's objectives, procedures, and potential risks. Participants were also informed of their right to withdraw at any time without consequence. To maintain confidentiality, all personal identifiers were removed before analysis.

The study was approved by the Ethics Committee of the Faculty of Education and Sports Sciences at the University of Vigo (Spain) (approval code: 3-0406-14).

Despite employing a rigorous methodology, this study has several limitations. The sample size (146 participants), while adequate for analysis, may not fully represent the broader older adult population, particularly in different geographic regions or among individuals with varying health conditions. The dataset was derived from a single location, which may not capture variations in fall risk due to cultural, environmental, or lifestyle differences, such as urban versus rural living or access to healthcare.

Additionally, reliance on wearable sensors for accelerometric data introduces potential biases related to user compliance and variability in sensor placement. Some participants may forget to wear the device, position it incorrectly, or remove it due to discomfort, which could affect data consistency.

Additionally, the age distribution and health conditions of the sample may not fully capture the heterogeneity of older adults globally. Populations with higher rates of chronic illnesses, differing physical activity levels, or varying nutritional statuses could exhibit distinct fall risk patterns. Future studies should include larger and more geographically diverse samples to validate these models across broader populations and ensure their applicability.

To mitigate this limitation, we employed robust preprocessing and model evaluation techniques, including cross-validation, to enhance result reliability despite the sample size. However, further research is necessary to confirm these findings in large-scale, multi-site studies.

Another limitation is the cross-sectional nature of the data, which assesses fall risk at a single point in time. Longitudinal studies are needed to validate the models over time and assess their effectiveness in real-world settings.

While integrating accelerometric and non-accelerometric data improves prediction accuracy, the complexity of these models may limit their clinical applicability, particularly in settings with limited computational resources.

Future research should address these limitations by incorporating larger, more diverse populations and exploring the application of these models in longitudinal and real-world scenarios.

## Experiments and results

### Dataset

The dataset used in this study was collected from a group of older adults, focusing on individuals aged 65 years and above. The data collection spanned over a period of 12 months and included both accelerometric and non-accelerometric variables. These variables were meticulously recorded to ensure a comprehensive dataset that captures the multidimensional aspects of fall risk. In total, the dataset comprised data from 146 participants, which included accelerometric measures of movement and balance, as well as non-accelerometric variables such as demographic information and health status. The detailed description of the variables is provided in Table 1.

### Experimental setup

Prior to machine learning analysis, the dataset underwent a comprehensive preprocessing phase, which included:

- Normalization of continuous variables to ensure comparability across features.
- Encoding of categorical variables into a numerical format suitable for machine learning models.
- Handling of missing data through imputation methods to maintain data integrity.

Following preprocessing, the input variables were categorized into three groups: accelerometric, non-accelerometric, and a combination of both. This categorization enabled a systematic evaluation of how different data types impact model performance.

The machine learning models applied in this study included random forest, XGBoost, AdaBoost, LightGBM, SVR, decision trees, and Bayesian ridge regression. Each model was optimized using grid search with five-fold cross-validation, systematically testing multiple hyperparameter combinations to maximize predictive performance.

To ensure robust and reliable results, the final evaluation employed eight-fold cross-validation, where the data were divided into eight subsets. Models were iteratively trained on seven subsets and tested on the remaining subset, minimizing overfitting and improving generalizability to new data.

### Results

The performance of the machine learning models was evaluated using three key metrics: MSE, MAE, and the coefficient of determination ($R^2$). These metrics were computed

**Table 2.** MSE, $R^2$, and MAE with standard deviations across cross-validation folds.

| Model | MSE (Std. Dev.) | $R^2$ (Std. Dev.) | MAE (Std. Dev.) |
| --- | --- | --- | --- |
| Random forest | 0.0608 (0.0059) | 0.9355 (0.0949) | 0.2096 (0.0204) |
| XGBoost | 0.0267 (0.0027) | 0.9703 (0.0977) | 0.1389 (0.0142) |
| LightGBM | 0.3527 (0.0350) | 0.7453 (0.0748) | 0.5048 (0.0501) |
| Bayesian ridge | 0.6746 (0.0678) | 0.9941 (0.1006) | 0.6981 (0.0685) |
| SVR | 1.5243 (0.1535) | −2.2532 (0.2175) | 1.0494 (0.0994) |
| Decision trees | 0.1907 (0.0187) | 0.8991 (0.0896) | 0.3712 (0.0377) |
| AdaBoost | 0.1282 (0.0127) | 0.8912 (0.0901) | 0.3043 (0.0309) |

MSE: mean squared error; $R^2$: coefficient of determination; MAE: mean absolute error; XGBoost: extreme gradient boosting; LightGBM: light gradient boosting machine; SVR: support vector regression; AdaBoost: adaptive boosting.

for each model and across three groups of variables: accelerometric, non-accelerometric, and a combination of both.

MSE was used to measure overall prediction error, MAE provided insight into the average absolute deviation from actual values, and $R^2$ assessed the proportion of variance explained by the models.

*Average performance by machine learning model.* Table 2 presents the average MSE, MAE, and $R^2$ for each machine learning model tested in this study. This table provides a comprehensive comparison of model performance across all configurations, highlighting which algorithms achieve the best predictive accuracy (MSE), explanatory power ($R^2$), and error robustness (MAE), regardless of the dataset type.

To complement this numerical analysis, Figures 2 and 3 visualize the results, incorporating CIs for a more complete assessment of model stability. These figures enable a clearer interpretation of how each model performs in relation to different groups of variables: accelerometric, non-accelerometric, and combined data.

Figure 2 compares MSE, $R^2$, and MAE across models, now including CIs to assess variability across cross-validation folds. The results indicate that XGBoost achieved the lowest MSE (0.0267 (0.0027)) and MAE (0.1389 (0.0142)), signifying its high predictive accuracy and MAE.

Conversely, SVR exhibited the highest MSE (1.5243 (0.1535)) and MAE (1.0494 (0.0994)), alongside a negative $R^2$ (−2.2532 (0.2175)), confirming its poor suitability for this task. Bayesian ridge regression attained the highest $R^2$ (0.9941 (0.1006)), suggesting superior explanatory power within this dataset, despite exhibiting higher MAE than tree-based models.

These findings reinforce the importance of considering not only predictive accuracy (MSE) and explanatory power ($R^2$) but also absolute error (MAE) when evaluating model reliability, as models with lower MAE provide more stable predictions with fewer large deviations.

*Average performance by group of variables.* Table 3 presents the average MSE, MAE, and $R^2$ for each group of variables: accelerometric, non-accelerometric, and combined. This table analyzes how different data sources influence overall model accuracy and explanatory power, independent of the specific machine learning algorithm used.

Unlike Table 2, which compares model performance, Table 3 highlights the predictive power inherent to each data group. The results indicate that the non-accelerometric group achieved the lowest MSE (188.98 (19.49)), the highest $R^2$ (0.98 (0.10)), and the lowest MAE (11.68 (1.15)), making it the most reliable predictor of fall risk.

In contrast, the accelerometric group exhibited the highest MSE (6823.99 (677.79)) and the highest MAE (70.22 (7.27)), alongside a negative $R^2$ (−0.16 (0.02)), confirming its poor predictive capability when used alone. The combined dataset demonstrated strong performance across all metrics, reinforcing that integrating both data types enhances predictive accuracy and model reliability.

The performance of the models was also analyzed based on the type of input variables used. Table 3 presents the numerical results, while Figure 3 provides a comparative visualization of MSE, $R^2$, and MAE across the three groups of variables. CIs are included to ensure a more robust analysis of predictive stability.

The results confirm that the accelerometric dataset exhibited the highest MSE (6823.99 (677.79)) and MAE (70.22 (7.27)), reinforcing its limited predictive capability when used alone. In contrast, the non-accelerometric dataset achieved the lowest MSE (188.98 (19.49)) and the highest $R^2$ (0.98 (0.10)), highlighting the importance of demographic and clinical variables in fall risk prediction.

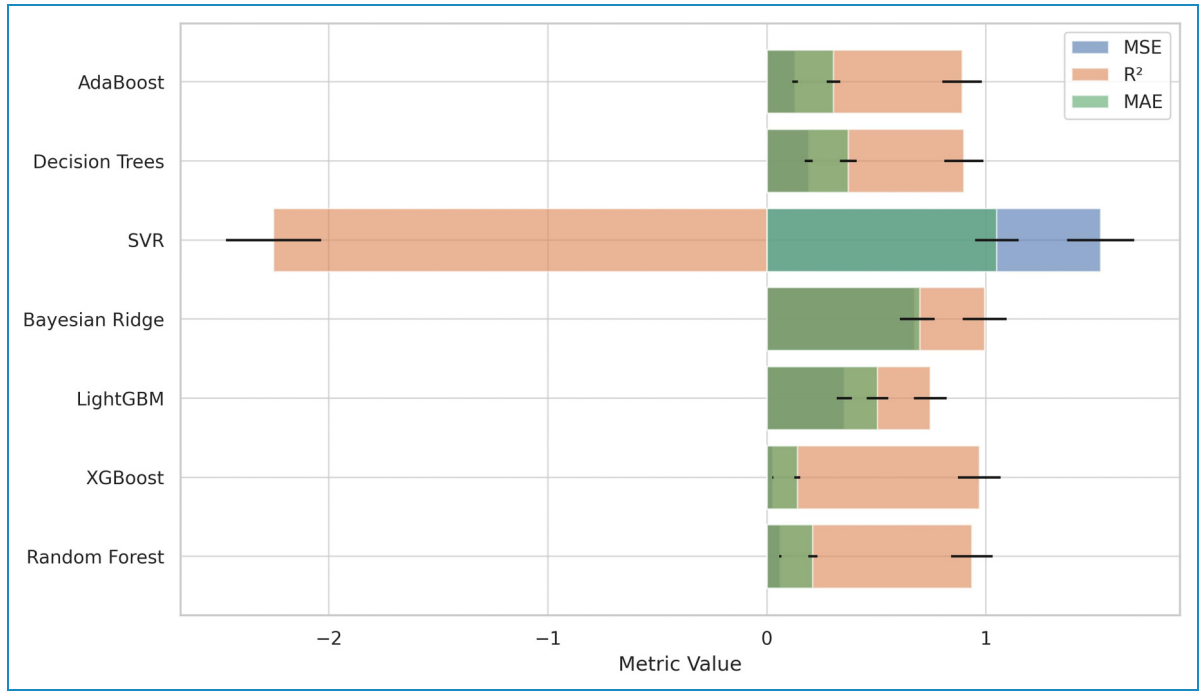

**Figure 2.** Model performance: MSE, $R^2$, and MAE with confidence intervals. MSE: mean squared error; $R^2$: coefficient of determination; MAE: mean absolute error.

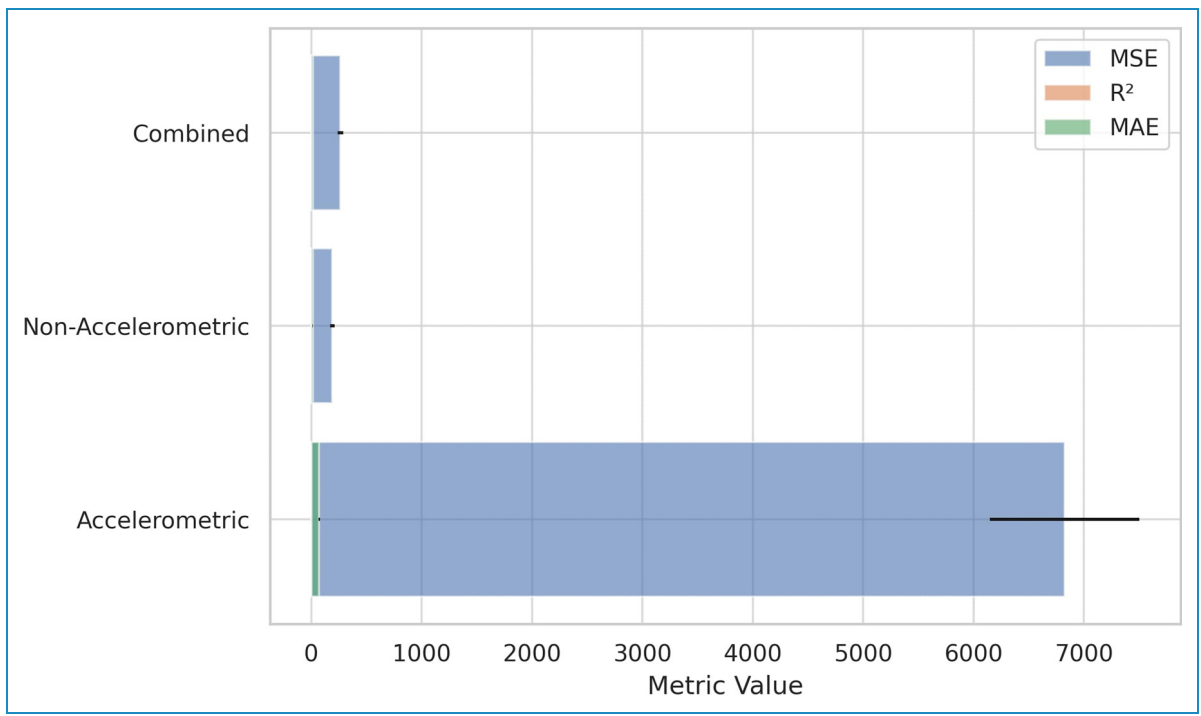

**Figure 3.** Group performance: MSE, $R^2$, and MAE with confidence intervals. MSE: mean squared error; $R^2$: coefficient of determination; MAE: mean absolute error.

The combined dataset demonstrated strong performance across all metrics, with MSE (262.25 (24.70)), MAE (13.77 (1.37)), and $R^2$ (0.97 (0.10)), validating that integrating multiple data types improves both predictive accuracy and explanatory power. Including CIs in the visualization provides a clearer assessment of the stability of each dataset's predictive performance across cross-validation folds.

Table 3. MSE with 95% CIs, $R^2$, and MAE with standard deviations for different groups of variables.

| Group of variables | MSE (Std. Dev.) | $R^2$ (Std. Dev.) | MAE (Std. Dev.) |
| --- | --- | --- | --- |
| Accelerometric | 6823.99 (677.79) | −0.16 (0.02) | 70.22 (7.27) |
| Non-accelerometric | 188.98 (19.49) | 0.98 (0.10) | 11.68 (1.15) |
| Combined | 262.25 (24.70) | 0.97 (0.10) | 13.77 (1.37) |

MSE: mean squared error; CIs: confidence intervals; $R^2$: coefficient of determination; MAE: mean absolute error.

## Discussion

The findings of this study underscore the critical importance of integrating both accelerometric and non-accelerometric data to enhance the predictive accuracy of machine learning models for fall risk assessment in older adults. Our results align with previous research highlighting the multifactorial nature of fall risk and the advantages of combining diverse data sources.[14,15]

Specifically, this study confirms that non-accelerometric data, such as demographic and health-related variables, significantly improve predictive performance, as evidenced by the lower MSE, lower MAE, and higher $R^2$ compared to models relying solely on accelerometric data.

The XGBoost model emerged as the most effective algorithm in terms of predictive accuracy, achieving the lowest MSE (0.0267 (0.0027)) and MAE (0.1389 (0.0142)) across different data types. However, Bayesian Ridge Regression attained the highest $R^2$ (0.9941 (0.1006)), indicating its strong explanatory power.

Conversely, SVR exhibited the highest MSE (1.5243 (0.1535)) and MAE (1.0494 (0.0994)), with a negative $R^2$ (−2.2532 (0.2175)), confirming its poor suitability for this task.

These findings suggest that ensemble models like XGBoost provide a balanced trade-off between accuracy and computational efficiency, while Bayesian ridge is particularly effective in explaining variance in fall risk predictions. The results reinforce the benefit of integrating diverse data types to enhance model robustness and clinical applicability.

The strong performance of Bayesian ridge regression in this study is likely due to its ability to effectively manage high-dimensional datasets with relatively small sample sizes. Unlike tree-based models such as LightGBM, which rely on splitting criteria that may be unstable in small datasets, Bayesian Ridge applies probabilistic regularization to prevent overfitting while capturing linear dependencies between variables.

This is particularly relevant in our study, where the integration of accelerometric and non-accelerometric features may introduce multicollinearity, a challenge that Bayesian ridge naturally handles.

In contrast, the lower performance of LightGBM compared to Bayesian ridge suggests that boosting-based models may require larger datasets to fully exploit complex feature interactions. While LightGBM excels in structured data analysis, its reliance on feature splitting may have led to overfitting or suboptimal feature selection given the dataset size.

These findings align with prior research showing that Bayesian methods outperform tree-based models in small-to-medium datasets with a high number of correlated variables.[22]

Interestingly, the results reveal that while accelerometric data alone provide valuable insights into movement-related risks, their predictive power is significantly limited, as indicated by the negative $R^2$ value (−2.2532 (0.2175)). However, when combined with non-accelerometric factors, predictive accuracy notably increases, resulting in one of the highest $R^2$ values (0.9941 (0.1006)).

This finding underscores the importance of contextual data, such as chronic health conditions, for a comprehensive assessment of fall risk. It supports the notion that fall risk is driven by a complex interplay of intrinsic and extrinsic factors, which cannot be fully captured through movement data alone.[12]

Among the accelerometric variables analyzed, the inclusion of the maximum RMS value aligns with previous research.[30,31] Additionally, the model incorporated vertical and medio-lateral axis variables, consistent with studies that examined gait patterns (although not exclusively in healthy subjects).[31–33]

The vertical and medio-lateral axes appear to be more sensitive than the transverse axis—which exhibits lower amplitude oscillations—for diagnosing balance and postural control impairments.[33,34] Medio-lateral axis movements are often associated with instabilities or compensations, potentially linked to age-related reductions in lower limb mobility, excess fat mass (particularly around the trunk), or degenerative and cardiovascular conditions.[35]

These findings reinforce the relationship between accelerometric and non-accelerometric variables, as previously observed in the literature.[36,37] Regarding the sagittal plane, gait accelerations may reflect muscular compensations due to loss of stability during the single-limb support phase.[38] However, it is important to note that other studies on accelerometry and balance have reported significant results across all three movement axes, supporting the broader utility of accelerometric analysis in fall risk prediction.[36,37]

The non-accelerometric dataset, when used in isolation, demonstrated strong predictive capabilities, achieving an $R^2$ (0.98) close to that of the combined dataset (0.97). This underscores the importance of factors beyond movement patterns in fall risk assessment. Variables such as age and comorbidities were particularly influential, reinforcing their critical role in fall risk prediction among older adults. These findings suggest that in contexts where accelerometric data are unavailable, non-accelerometric information can still provide a robust predictive foundation.

Among the non-accelerometric variables, several emerged as key contributors to fall risk prediction. Age is widely recognized as a major factor, as advancing age correlates with declines in physical function, balance, and overall mobility. Similarly, comorbidities such as diabetes and cardiovascular diseases impair sensory-motor control, increasing fall susceptibility. Body mass index (BMI) also plays a dual role, where both underweight and obesity contribute to balance impairments, either due to reduced muscle strength or altered biomechanics.

These variables not only affect individual fall risk but also interact with accelerometric factors, providing a more nuanced understanding of risk. For instance, a high BMI may exacerbate gait instability detected by accelerometric sensors, while certain comorbidities could amplify movement irregularities, further influencing fall likelihood. By incorporating these factors, non-accelerometric data enhance the models' ability to identify at-risk individuals with higher specificity.

The inclusion of anthropometric variables is consistent with previous findings on their impact on balance and postural control.[36,39] Specifically, BMI and trunk circumference are known to affect static and dynamic balance.[39,40] Bioimpedance measures, which reflect body composition and fat percentage, have also been linked to postural deviations in the antero-posterior axis, negatively affecting stability.[39,41]

Similarly, existing research has highlighted the influence of cardiovascular and degenerative conditions on fall risk.[42–44] Arrhythmias, syncope, and hypotension are particularly associated with an increased likelihood of falls.[43] In the case of degenerative conditions, one consequence is scapular and pelvic girdle misalignment, which directly impacts postural control and further elevates fall risk.[44,45]

Finally, several studies confirm the predictive value of functional tests—such as those used in this research—in assessing fall risk.[46–48] However, there is still no consensus on the single most predictive functional test.[48]

However, it is important to note that the SVR model, despite its widespread use in machine learning tasks, performed the worst across all datasets. This is reflected in its high MSE (1.5243 (0.1535)), high MAE (1.0494 (0.0994)), and negative $R^2$ (−2.2532 (0.2175)). These results suggest that SVR struggled to generalize effectively, likely due to its sensitivity to parameter tuning and the nature of the dataset used.

Several factors contributed to the poor performance of SVR in this study.

First, SVR is highly sensitive to feature scaling and kernel selection, requiring extensive fine-tuning to achieve optimal performance. Given that our dataset integrates both accelerometric and non-accelerometric features, with varying magnitudes and distributions, SVR likely struggled to balance these differences effectively.

Second, SVR's computational complexity increases significantly with the number of features, making it less scalable in high-dimensional datasets like ours. Unlike Bayesian ridge, which effectively handles multicollinearity through regularization, SVR depends on careful kernel function selection to model complex relationships. If the chosen kernel fails to capture the underlying patterns, the model fails to generalize, leading to the high error rates and negative $R^2$ values observed.

These findings align with prior research, which has highlighted SVR's limitations in datasets with mixed data types and high feature dimensionality, reinforcing the need for careful model selection based on dataset characteristics.[25]

The implications of these findings are significant for the development of fall prevention strategies. By integrating both accelerometric and non-accelerometric data, healthcare providers can develop more accurate and personalized fall risk assessments. This, in turn, enables targeted interventions, reducing fall incidence and improving quality of life for older adults.

Moreover, the strong performance of Bayesian ridge regression suggests that future research should further explore Bayesian methods in clinical prediction, particularly in the context of multidimensional risk factors.

To further validate the robustness of our findings, we examined the variability of model performance across cross-validation folds. In addition to reporting average MSE, $R^2$, and MAE, we now include standard deviations for each metric, providing insight into the consistency of model predictions.

The results indicate that Bayesian ridge not only achieved the highest $R^2$ (0.9941 (0.1006)) but also exhibited the lowest variability across folds, reinforcing its stability in high-dimensional but small datasets.

In contrast, tree-based models like LightGBM and XGBoost showed greater fluctuations, likely due to their reliance on iterative feature selection, which can lead to performance variations depending on the training subset. The highest variability was observed in SVR, further confirming its difficulty in handling heterogeneous data, as reflected by its MSE (1.5243 (0.1535)), MAE (1.0494 (0.0994)), and negative $R^2$ (−2.2532 (0.2175)).

These additional analyses strengthen the reliability of our findings, ensuring that model performance is not solely dependent on specific training-test splits but remains consistent across different cross-validation iterations. The standard deviations for each model have been incorporated into Tables 2 and 3.

In line with existing literature, our study reinforces the potential of machine learning to advance fall risk prediction.[12,15] By leveraging diverse data sources, machine learning models provide a more nuanced understanding of fall risk factors, enabling better-informed clinical decisions.

However, further research is needed to assess the generalizability of these findings across different populations and clinical settings, as well as to refine predictive models for even greater accuracy.

In conclusion, this study contributes to the growing body of evidence supporting the integration of accelerometric and non-accelerometric data in fall risk prediction. This combination not only enhances predictive performance but also provides a holistic view of the risk factors involved.

## Comparison with previous studies

The findings of this study align with and expand upon recent research on fall risk prediction among older adults. For example, Antonietti[18] and Millet et al.[19] demonstrated the utility of accelerometric data in predicting fall risk, particularly for gait stability and movement patterns.

However, unlike these studies, our results highlight the significant predictive power of non-accelerometric variables, such as age, comorbidities, and BMI, especially when combined with accelerometric data. This integration resulted in higher predictive accuracy and explanatory power, achieving an $R^2$ of 0.97 with combined data, compared to models relying solely on accelerometric variables.

In contrast to Urbanek et al.,[11] who focused exclusively on accelerometry-based assessments, our study demonstrates that integrating non-accelerometric data enhances risk factor detection, capturing factors that movement-based metrics alone may overlook. Specifically, incorporating comorbidities and demographic data improved our model's ability to identify individuals at high fall risk with greater sensitivity.

Additionally, while Sharma et al.[20] and Sasso et al.[21] employed advanced machine learning models, their analyses were limited to either accelerometric or clinical datasets in isolation.

Our approach differs by systematically comparing the predictive value of accelerometric, non-accelerometric, and combined datasets, offering a more comprehensive evaluation of their relative contributions to fall risk prediction.

These comparisons highlight the novelty of this work in integrating diverse data sources and leveraging machine learning models to achieve higher predictive accuracy. By addressing key gaps in previous studies, such as the lack of data integration and limited dataset generalizability, our findings contribute to the advancement of fall risk prediction in older adults.

Future research should focus on:

- Validating these models in larger and more diverse populations to ensure generalizability.
- Exploring other machine learning algorithms, including hybrid models combining deep learning with structured ensemble approaches.
- Investigating the clinical application of these findings to ensure that predictive models translate into real-world improvements in fall prevention and patient care.

## Conclusions

This study provides valuable insights into the application of machine learning techniques for predicting fall risk among older adults, emphasizing the importance of integrating both accelerometric and non-accelerometric data.

Our findings demonstrate that models incorporating non-accelerometric factors, such as age and comorbidities, significantly enhance predictive accuracy compared to models relying solely on accelerometric data, which exhibited the poorest performance in terms of both MSE (1.5243 (0.1535)) and $R^2$ (−2.2532 (0.2175)).

The XGBoost model emerged as the most effective in terms of predictive accuracy, achieving the lowest MSE (0.0267 (0.0027)), while Bayesian ridge attained the highest $R^2$ (0.9941 (0.1006)), demonstrating its superior ability to explain variance in fall risk predictions across different data configurations.

Despite these promising results, several limitations should be acknowledged.

- The dataset, although comprehensive, included only 146 participants, which may limit the generalizability of the findings.
- The observational study design and specific population sample could introduce biases affecting predictive model performance in different settings.
- Certain risk factors not captured in the dataset, such as detailed health history beyond the included variables, may influence fall risk. Incorporating these additional factors in future research could enhance predictive accuracy and provide a more nuanced understanding of fall risk.

Nevertheless, this study underscores the potential of combining accelerometric and non-accelerometric data to develop robust predictive models for fall risk. The strong performance of both XGBoost and Bayesian ridge highlights the importance of using advanced statistical and machine learning methods to manage complex, multidimensional datasets commonly found in clinical research.

These findings contribute to the growing evidence supporting the use of machine learning in fall prevention strategies, offering a more comprehensive approach to risk assessment.

In conclusion, while further validation is necessary, this study advances our understanding of how different data types contribute to fall risk prediction in older adults. Future studies should focus on:

- Expanding the dataset to improve generalizability.
- Exploring additional variables that may further enhance predictive performance.
- Refining machine learning models to improve accuracy and clinical applicability.

Additionally, incorporating advanced model interpretability techniques will be essential to better understand the contributions of individual variables in fall risk prediction. Given the high-dimensional nature of our dataset (337 variables) and the limited sample size (146 participants), feature importance estimates may be highly variable and sensitive to minor changes in the training set.

For this reason, this study did not include explicit feature importance analysis. However, applying SHAP (SHapley Additive Explanations) or permutation importance in future research will enhance model transparency, facilitating clinical adoption and enabling the development of more effective, personalized interventions aimed at reducing fall-related incidents and improving the quality of life for older adults.

## Guarantor
José Alberto Benítez-Andrades.

## ORCID iDs

Ana González-Castro  https://orcid.org/0000-0001-7999-4017
José Alberto Benítez-Andrades  https://orcid.org/0000-0002-4450-349X
Camino Prada-García  https://orcid.org/0000-0002-2487-9350

## Supplemental material
Supplemental material for this article is available online.

## Statements and declarations

### Ethical considerations
According to the protocol approved by the Ethics Committee for Research with Medicines of the Universidad de León. Informed consent is not required for this study.

### Author contributions/CRediT
Ana González-Castro: data curation, methodology, writing–original draft preparation, and writing–reviewing and editing. José Alberto Benítez-Andrades: conceptualization, data curation, methodology, software, visualization, validation, supervision, writing–original draft preparation, and writing–reviewing and editing. Camino Prada-García: conceptualization, visualization, validation, and writing-original draft preparation. Rubén González-González: data curation, software, visualization, and writing–reviewing and editing. Raquel Leirós-Rodríguez: conceptualization, data curation, methodology, validation, supervision, and writing–original draft preparation.

### Funding
The author(s) disclosed receipt of the following financial support for the research, authorship, and/or publication of this article: This work is a result of the project "NLP-Driven Insight Engine for Suicide Attempt Detection in EHR (SUICIDETECT)", that is being developed under grant "PID2023-146168OA-I00" from the Spanish Ministerio de Ciencia, Innovación y Universidades.

### Conflicting interests
The author(s) declared no potential conflicts of interest with respect to the research, authorship, and/or publication of this article.